\newcommand{\autodevCodeGenPassOne}{91.5}
\newcommand{\humanTestGenCoverage}{99.4}
\newcommand{\autodevTestGenPassOne}{87.8}
\newcommand{\autodevTestGenCoverage}{99.3}

\documentclass[sigplan,screen, authorversion, nonacm]{acmart}
\AtBeginDocument{%
  \providecommand\BibTeX{{%
    \normalfont B\kern-0.5em{\scshape i\kern-0.25em b}\kern-0.8em\TeX}}}
\usepackage[frozencache,cachedir=.]{minted}
\setminted{fontsize=\scriptsize}
\usepackage{verbatim}
\usepackage{xcolor}
\usepackage{multirow}
\usepackage{tcolorbox}
\usepackage{booktabs}
\usepackage{tabularx}
\usepackage{adjustbox}
\usepackage{listings}
\DeclareCaptionStyle{ruled}{labelfont=normalfont,labelsep=colon,strut=off} 
\floatstyle{ruled}
\newfloat{listing}{tb}{lst}{}
\floatname{listing}{Listing}

\begin{document}

\title{AutoDev: Automated AI-Driven Development}





\author{Michele Tufano}
\affiliation{%
\institution{Microsoft}
  \state{Redmond}
  \country{USA}
}

\author{Anisha Agarwal}
\affiliation{%
\institution{Microsoft}
  \state{Redmond}
  \country{USA}
}

\author{Jinu Jang}
\affiliation{%
\institution{Microsoft}
  \state{Redmond}
  \country{USA}
}

\author{Roshanak Zilouchian Moghaddam}
\affiliation{%
\institution{Microsoft}
  \state{Redmond}
  \country{USA}
}

\author{Neel Sundaresan}
\affiliation{%
\institution{Microsoft}
  \state{Redmond}
  \country{USA}
}

\renewcommand{\shortauthors}{Michele Tufano, et al.}

\begin{abstract}
The landscape of software development has witnessed a paradigm shift with the advent of AI-powered assistants, exemplified by GitHub Copilot. However, existing solutions are not leveraging all the potential capabilities available in an IDE such as building, testing, executing code, git operations, etc. Therefore, they are constrained by their limited capabilities, primarily focusing on suggesting code snippets and file manipulation within a chat-based interface.

To fill this gap, we present AutoDev, a fully automated AI-driven software development framework, designed for \emph{autonomous} planning and execution of intricate software engineering tasks. AutoDev enables users to define complex software engineering objectives, which are assigned to AutoDev's autonomous AI Agents to achieve. These AI agents can perform diverse operations on a codebase, including file editing, retrieval, build processes, execution, testing, and git operations. They also have access to files, compiler output, build and testing logs, static analysis tools, and more. This enables the AI Agents to execute tasks in a fully automated manner with a comprehensive understanding of the contextual information required. Furthermore, AutoDev establishes a secure development environment by confining all operations within Docker containers. This framework incorporates guardrails to ensure user privacy and file security, allowing users to define specific permitted or restricted commands and operations within AutoDev. 

In our evaluation, we tested AutoDev on the HumanEval dataset, obtaining promising results with \autodevCodeGenPassOne\% and \autodevTestGenPassOne\% of Pass@1 for code generation and test generation respectively, demonstrating its effectiveness in automating software engineering tasks while maintaining a secure and user-controlled development environment.



\end{abstract}

\maketitle

\definecolor{ForestGreen}{RGB}{34,139,34}
\definecolor{RoyalBlue}{RGB}{85,118,209}
\definecolor{DarkPink}{RGB}{255,110,236}

\definecolor{Gray}{gray}{0.9}
\newcommand{\txtmint}[1]{\mintinline[fontsize=\scriptsize, bgcolor=Gray]{text}{#1}}

\newcommand{\dataset}{{\sc CoverageEval}\xspace}

\newcommand{\ie}{\textit{i.e.,}~}
\newcommand{\eg}{\textit{e.g.,}~}
\newcommand{\etc}{\textit{etc.}~}
\newcommand{\etal}{\textit{et al.}~}

\newcommand{\nb}[2]{
    \fbox{\bfseries\sffamily\scriptsize#1}
    {\sf\small$\blacktriangleright$\textit{#2}$\blacktriangleleft$}
}

\newcommand\MICHELE[1]{\textcolor{blue}{\nb{MICHELE}{#1}}}
\newcommand\ANISHA[1]{\textcolor{ForestGreen}{\nb{ANISHA}{#1}}}
\newcommand\ROSHANAK[1]{\textcolor{DarkPink}{\nb{ROSHANAK}{#1}}}

\begin{figure}[ht]
    \centering
    \includegraphics[width=0.5\textwidth]{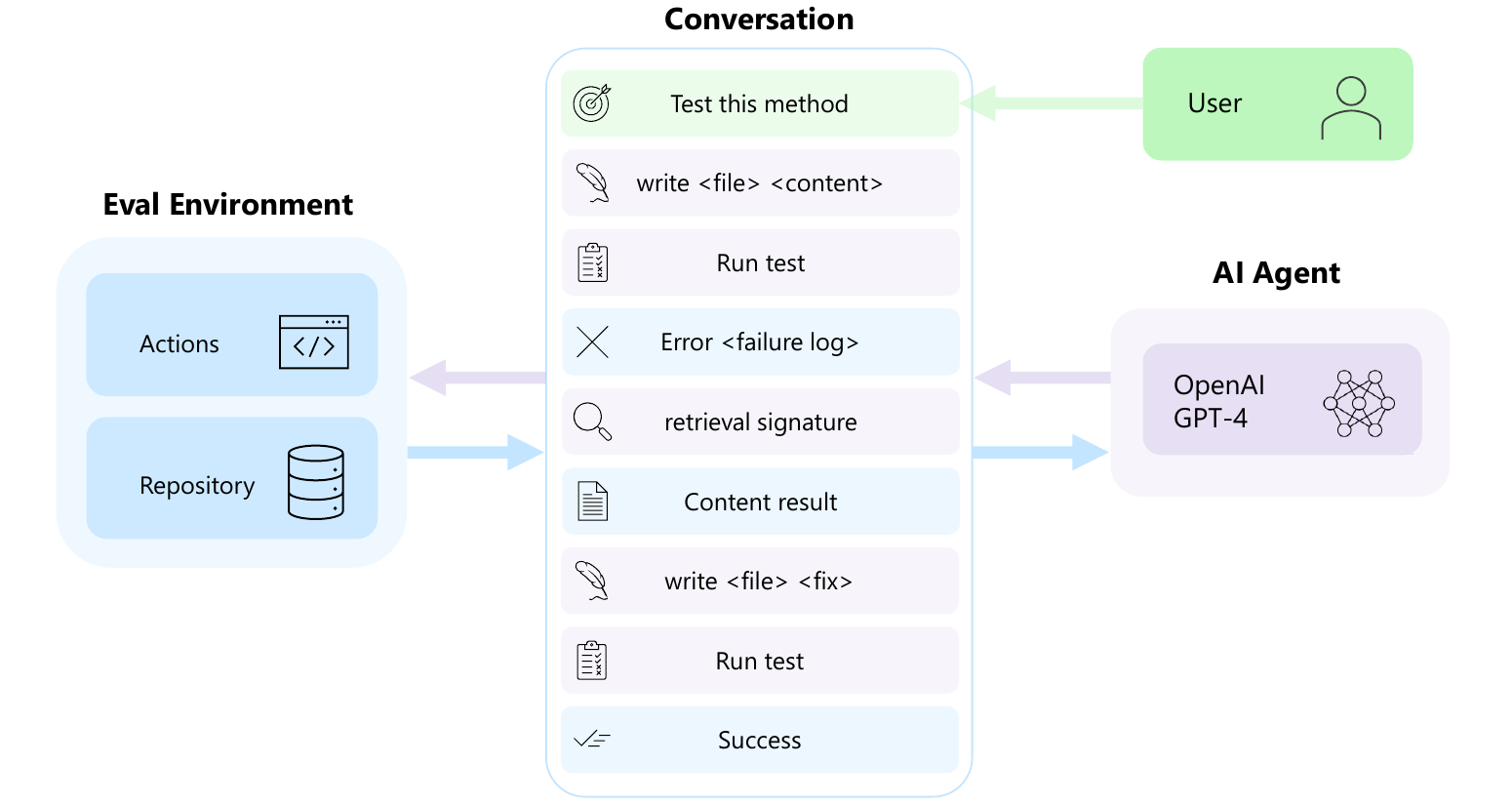}
    \caption{AutoDev enables an AI Agent to achieve a given objective by performing several actions within the repository. The Eval Environment executes the suggested operations, providing the AI Agent with the resulting outcome. In the conversation, purple messages are from the AI agent, while blue messages are responses from the Eval Environment.}
    \label{fig:overview}
\end{figure}

\section{Introduction}
As developers increasingly adopt AI assistants such as ChatGPT for their development tasks, productivity gains become evident. AI coding assistants have further advanced into integrated development environments (IDEs) like GitHub Copilot \cite{copilot}, where they offer code suggestions both within chat interfaces and directly within files.

However, these AI coding assistants, despite their integration into IDEs, exhibit limited functionalities and lack contextual awareness\cite{ding2024crosscodeeval, agarwal2024copilot}. They often do not leverage all IDE capabilities such as invoking linters, compilers, or executing command-line operations, and consequently, developers still need to manually validate syntax and ensure the correctness of AI-generated code, execute the codebase, and inspect error logs.

AutoDev bridges this gap by offering autonomous AI agents the ability to execute actions such as file editing, retrieval, build, testing, and CLI commands directly within the repository to achieve user-defined objectives, thus enabling the completion of complex tasks autonomously.

AutoDev offers the following key features: (i) the ability to track and manage user and AI agents conversations through a \emph{Conversation Manager}, (ii) a library of customized \emph{Tools} to accomplish a variety of code and SE related objectives, (iii) the ability to schedule various AI agents to work collaboratively towards a common objective through an \emph{Agent Scheduler}, and (iv) the ability to execute code and run tests through an \emph{Evaluation Environment}.   

Figure \ref{fig:overview} illustrates a high-level example of the AutoDev workflow. The user defines an objective (e.g., testing a specific method). The AI Agent writes tests in a new file and initiates the test execution command, all within a secure Evaluation Environment. The output of the test execution, including failure logs, is then incorporated into the conversation. The AI agent analyzes this output, triggers a retrieval command, incorporates the retrieved information by editing the file, and re-invokes the test execution. Finally, the environment provides feedback on the success of the test execution and completion of the user's objective.

The entire process is orchestrated by AutoDev autonomously, requiring no developer intervention beyond setting the initial objective. In contrast, with existing AI coding assistants integrated into IDEs, developers would have to manually execute tests (e.g., run pytest), provide failure logs to the AI chat interface, possibly identify additional contextual information to be incorporated, and repeat validation actions to ensure test success after the AI generates revised code.


AutoDev draws inspiration from previous work in the area of autonomous AI agents. For example, AutoGen \cite{wu2023autogen} is a framework which orchestrates language model workflows and facilitates conversations between multiple agents. AutoDev extends AutoGen by going beyond conversation management and enabling agents to directly interact with the code repository, executing commands and actions autonomously. Similarly, AutoDev builds upon Auto-GPT \cite{autogpt}, an open-source AI agent for autonomous task execution by offering code and IDE specific capabilities to enable execution of complex software engineering tasks. 

In our evaluation, we assess the capabilities of AutoDev using the HumanEval dataset\cite{chen2021codex}, originally designed for code generation from natural language descriptions (docstrings). Additionally, we extend the evaluation to include the test case generation task, showcasing the versatility of AutoDev in handling diverse software engineering objectives. The results demonstrate promising performance, with AutoDev achieving remarkable scores of \autodevCodeGenPassOne\% and \autodevTestGenPassOne\% for Pass@1 for code generation and test generation respectively. These outcomes underscore the effectiveness of AutoDev in automating software engineering tasks while maintaining a secure and user-controlled development environment.

\begin{figure*}[h]
    \centering
    \includegraphics[width=1.0\textwidth]{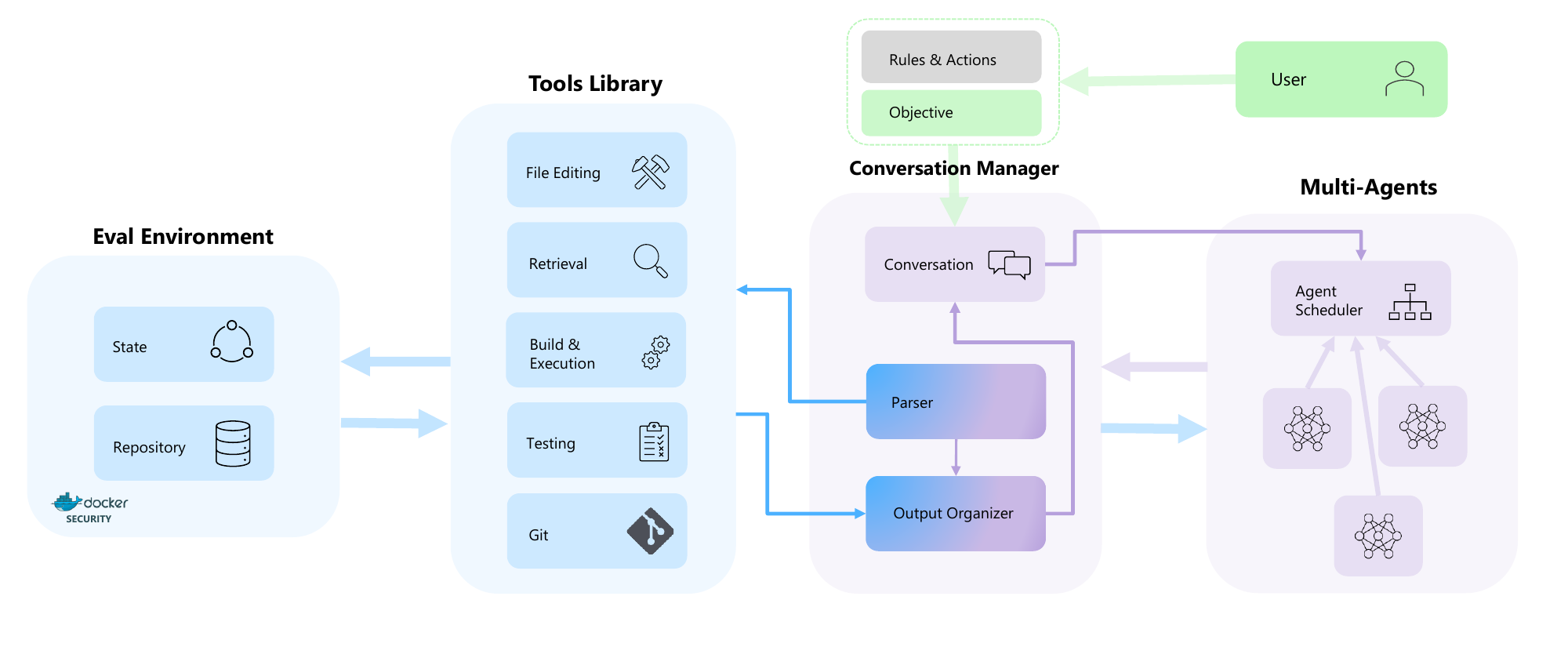}
    \caption{Overview of the AutoDev Framework: The user initiates the process by defining the objective to be achieved. The Conversation Manager initializes the conversation and settings. The Agent Scheduler orchestrates AI agents to collaborate on the task and forwards their commands to the Conversation Manager. The Conversation Manager parses these commands and invokes the Tools Library, which offers various actions that can be performed on the repository. Agents' actions are executed within a secure Docker environment, and the output is returned to the Conversation Manager, which incorporates it into the ongoing conversation. This iterative process continues until the task is successfully completed.}
    \label{fig:design}
\end{figure*}

\section{AutoDev Design}

The design overview depicted in Fig. \ref{fig:design} provides insight into AutoDev's architecture. Once the initial configurations are complete, AutoDev organizes its capabilities into four groups: a Conversation Manager that tracks and manages the user and agents conversations; a Tools library where a variety of code and IDE related tools are available for the agents, an Agents Scheduler that schedules various agents, and an
Evaluation Environment that enables execution operations. Below we explain each capability in details.
\subsection{Rules, Actions, and Objective Configuration}
The user initiates the process by configuring rules and actions through yaml files. These files define the available commands (actions) that AI agents can perform. Users can leverage default settings or fine-grained permissions by enabling/disabling specific commands, tailoring AutoDev to their specific needs. This configuration step allows for precise control over the AI agents' capabilities. At this stage the user can define the number and behavior of the AI agents, assigning specific responsibilities, permissions, and available actions. For example, the user could define a "Developer" agent and a "Reviewer" agent, that collaboratively work towards an objective.

Following the rules and actions configuration, the user specifies the software engineering task or process to be accomplished by AutoDev. For example, the user can ask to generate test cases and make sure they are syntactically correct, pass, and that do not contain bugs (this involves editing files, running test suite, executing syntax checks and bug-finding tools).

\subsection{Conversation Manager}
The Conversation Manager, responsible for initializing the conversation history, plays a pivotal role in overseeing the high-level management of the ongoing conversation. It handles the task of deciding when to interrupt the process and ensures seamless communication between the user, AI agents, and the system as a whole. It maintains a conversation object which includes the messages from the AI agents and the result of the actions from the Evaluation Environment.

\subsubsection{Parser}
The Parser interprets the responses generated by agents, extracting commands and arguments in a predefined format. It ensures that the instructions are correctly formatted, validating the number and accuracy of arguments (e.g., a file editing command requires the file path argument). In case of parsing failures, error messages are injected into the conversation, preventing further actions on the repository. Successfully parsed commands are further analyzed, by enforcing specific agent permissions and conducting additional semantic checks. It ensures that the suggested actions comply with the fine-grained permissions specified by the user. If the command passes scrutiny, the conversation manager invokes the corresponding action in the tools library.

\subsubsection{Output Organizer}
The Output Organizer module processes the output received from the Evaluation Environment. It selects crucial information, such as status or errors, optionally summarizes relevant content, and adds a well-structured message to the conversation history. This ensures that the user has a clear and organized record of the AutoDev's actions and outcomes.

\subsubsection{Conversation Conclusion}
The Conversation Manager determines when to conclude the conversation. This may happen when an agent signals the completion of the task (\texttt{stop} command), the conversation reaches a user-defined maximum number of iterations/tokens, or issues are detected either in the process or within the Evaluation Environment. The comprehensive design of AutoDev ensures a systematic and controlled approach to AI-driven development.

\subsection{Agent Scheduler}
The Agent Scheduler takes charge of orchestrating AI agents to achieve the user-defined objective. Agents, configured with specific personas and sets of available commands, operate collaboratively to perform diverse tasks. The scheduler employs various collaboration algorithms, such as Round Robin, Token-Based, or Priority-Based, to determine the order and manner in which agents contribute to the conversation. Specifically, scheduling algorithms include but not limited to: (i) Round Robin collaboration, which invokes each agent sequentially, allowing each to execute a predetermined number of operations; (ii) token-based collaboration, enabling an agent to undertake multiple operations until it issues a token signifying completion of its assigned tasks; (iii) priority-based collaboration, initiating agents in accordance with their prioritized order. The agent scheduler invokes a specific agent passing the current conversation.

\subsubsection{Agents}
Agents, comprising Large Language Models (LLMs) like OpenAI GPT-4 and Small Language Models (SLMs) optimized for code generation, communicate through textual natural language. These agents receive objectives and conversation histories from the Agent Scheduler, responding with actions specified by the Rules and Actions configuration. Each agent, with its unique configuration, contributes to the overall progress toward achieving the user's objective.

\subsection{Tools Library}
The Tools Library within AutoDev provides a range of commands that empower Agents to perform diverse operations on the repository. These commands are designed to encapsulate complex actions, tools, and utilities behind a straightforward and intuitive command structure. For instance, intricacies related to build and test execution are abstracted away through simple commands like \texttt{build} and \texttt{test <test\_file>}.

\begin{itemize}
    \item \textbf{File Editing}: This category encompasses commands for editing files, including code, configuration, and documentation. The utilities within this category, such as write, edit, insert, and delete, offer varying levels of granularity. Agents can perform actions ranging from writing entire files to modifying specific lines within a file. For example, the command \texttt{write <filepath> <start\_line>-<end\_line> <content>}, allows the agent to re-write a range of lines with new content.
    
    \item \textbf{Retrieval}: In this category, retrieval utilities range from basic CLI tools like grep, find, and ls to more sophisticated embedding-based techniques. These techniques enable Agents to look up similar code snippets, enhancing their ability to retrieve relevant information from the codebase. For example, the command \texttt{retrieve <content>} allows the agent to perform embedding-based retrieval of similar snippets to the provided content.

    \item \textbf{Build \& Execution}: Commands in this category allow Agents to compile, build, and execute the codebase effortlessly with simple and intuitive commands. The intricacies of low-level build commands are abstracted, streamlining the process within the Evaluation Environment infrastructure. Examples of commands in this category include: \texttt{build}, \texttt{run <file>}.

    \item \textbf{Testing \& Validation}: These commands enable Agents to test the codebase by executing a single test case, a specific test file, or the entire test suite. Agents can perform these actions without relying on low-level commands specific to particular testing frameworks. This category also encompasses validation tools such as linters and bug-finding utilities. Examples of commands in this category include: \texttt{syntax <file>} which checks the syntax correctness, and \texttt{test} which runs the entire test suite.

    \item \textbf{Git}: Fine-grained permissions for git operations can be configured by the user. This includes operations like commits, push, and merges. For instance, Agents can be granted permission to perform only local commits or, if necessary, push changes to the origin repository.

    \item \textbf{Communication}: Agents can invoke a set of commands aimed at facilitating communication with other agents and/or the user. Notably, the \texttt{talk} command enables sending natural language messages (not interpreted as commands for repository actions), the \texttt{ask} command is used to request user feedback, and the \texttt{stop} command interrupts the process, indicating goal achievement or the inability of the agents to proceed further.
\end{itemize}

The Tools Library in AutoDev thus provides a versatile and accessible set of tools for AI Agents to interact with the codebase and communicate effectively within the collaborative development environment.

\subsection{Evaluation Environment}
Running within a Docker container, the Evaluation Environment allows secure execution of file editing, retrieval, build, execution, and testing commands. It abstracts away the complexity of low-level commands, providing a simplified interface for agents. The  Evaluation Environment returns standard output/error to the Output Organizer module.

\subsection{Putting Everything Together}
The user initiates the conversation by specifying the objective and associated settings. The conversation manager initializes a conversation object, consolidating messages from both AI agents and the Evaluation Environment. Subsequently, the conversation manager dispatches the conversation to the Agent Scheduler, responsible for coordinating the actions of AI agents. In their role as AI agents, Language Models (Large or Small LMs) suggest commands through textual interactions.

The Commands Interface encompasses a diverse set of functionalities, including File Editing, Retrieval, Build and Execution, Testing, and Git operations. These suggested commands are then parsed by the Conversation Manager, which subsequently directs them to the Evaluation Environment for execution on the codebase.

Execution of these commands occurs within the secure confines of the Evaluation Environment, encapsulated within a Docker container. Post-execution, the resulting actions seamlessly integrate into the conversation history, contributing to subsequent iterations. This iterative process persists until the task is deemed complete by the Agents, user intervention occurs, or the maximum iteration limit is reached. AutoDev's design ensures a systematic and secure orchestration of AI agents to achieve complex software engineering tasks in an autonomous and user-controlled manner.

\section{Empirical Design}

In our empirical evaluation, we aim to assess AutoDev's capabilities and effectiveness in software engineering tasks, examining whether it can enhance an AI model's performance beyond simple inference. Additionally, we are interested in evaluating the cost of AutoDev in terms of the number of steps, inference calls, and tokens. We define three experimental research questions:

\subsection*{Research Questions}
\begin{enumerate}
    \item $RQ_1$: How effective is AutoDev in code generation task?
    \item $RQ_2$: How effective is AutoDev in test generation task?
    \item $RQ_3$: How efficient is AutoDev in completing tasks?
\end{enumerate}

\subsection*{$RQ_1$: How effective is AutoDev in code generation task?}

To address $RQ_1$, we evaluate AutoDev's performance in a code generation task using the HumanEval problem-solving dataset in Python. This dataset comprises 164 handwritten programming problems, each containing a function signature, docstring, body, and an average of 7.7 unit tests. In our assessment, AutoDev is provided with a partial file containing the function signature and docstring, with the objective of implementing the method.

We gauge AutoDev's effectiveness using the Pass@k metric, where \( k \) represents the number of attempts made. A successfully solved problem is defined as one where AutoDev generates the code of the method's body, satisfying all human-written tests. An attempt corresponds to an entire AutoDev conversation, which involves multiple inference calls and steps. This contrasts with other approaches, like directly invoking GPT-4, which typically involve a single inference call. Details regarding the multiple inference calls and steps are further explored in $RQ_3$. For this evaluation, we set \( k = 1 \), thus computing Pass@1, considering only the success achieved in the first attempt.

\subsection*{$RQ_2$: How effective is AutoDev in test generation task?}

For this research question, we modify the HumanEval dataset to evaluate AutoDev's capabilities in test generation. We consider the human-written solution and discard the provided human-written tests. AutoDev is instructed to generate test cases for the focal method and is evaluated based on test success, invocation of the focal method, and test coverage. We report Pass@1, considering tests successful if they pass and invoke the focal method. Additionally, we compare the coverage of AutoDev's tests with those written by humans.

\subsection*{$RQ_3$: How efficient is AutoDev in completing tasks?}

In this research question, we investigate AutoDev's efficiency in completing SE tasks. We analyze the number of steps or inference calls needed, the distribution of commands used (e.g., \texttt{write}, \texttt{test}), and the total number of tokens used in the conversation.

\subsection*{AutoDev Settings}

For this evaluation, AutoDev maintains consistent settings with one agent based on the GPT-4 model (gpt-4-1106-preview). Enabled actions include file editing, retrieval, and testing. The only communication command available is the \texttt{stop} command, indicating task completion. Other commands, such as \texttt{ask}, are disabled, requiring AutoDev to operate autonomously without human feedback or intervention beyond the initial goal-setting.

\section{Empirical Results}

\begin{table}[t]
\small 
\centering
\begin{tabular}{@{}lccc@{}}
\toprule
Approach                   & Model & Extra Training & Pass@1 \\ \midrule
Language Agent Tree Search & GPT-4 & \checkmark     & 94.4   \\
\textbf{AutoDev}           & GPT-4 & $\times$       & \autodevCodeGenPassOne   \\
Reflexion                  & GPT-4 & \checkmark     & 91.0   \\
zero-shot (baseline)       & GPT-4 & $\times$       & 67.0   \\ \bottomrule
\end{tabular}
\caption{Code Generation results on HumanEval. AutoDev achieves top-3 performance on the leaderboard without extra training data, unlike LATS and Reflexion.}
\label{tab:rq1}
\end{table}

\subsection*{$RQ_1$: How effective is AutoDev in a code generation task?}

Table \ref{tab:rq1} displays the results for $RQ_1$, comparing AutoDev against two alternative approaches and the zero-shot baseline. The table includes information about the model powering each approach, the need for additional training, and the Pass@1 metric.

We compared AutoDev with Language Agent Tree Search (LATS) and Reflexion, two leading approaches on the HumanEval leaderboard as of March 2024 \cite{paperswithcode-humaneval}. The results for the zero-shot baseline (GPT-4) are taken from the OpenAI GPT-4 technical report \cite{openai2024gpt4}, while those for LATS and Reflexion from the HuamnEval leaderboard \cite{paperswithcode-humaneval}.

Language Agent Tree Search (LATS) \cite{zhou2023language} is a versatile framework that utilizes Large Language Models (LLMs) for planning, acting, and reasoning. Inspired by Monte Carlo tree search, LATS employs LLMs as agents, value functions, and optimizers, repurposing their capabilities for improved decision-making.

Reflexion\cite{shinn2023reflexion} introduces a unique framework to reinforce language agents via linguistic feedback without weight updates. Agents in Reflexion verbally reflect on task feedback signals, maintaining their reflective text in an episodic memory buffer for enhanced decision-making.

Table \ref{tab:rq1} indicates that AutoDev achieves a Pass@1 rate of \autodevCodeGenPassOne\%, securing the second-best position on the HumanEval leaderboard. Notably, this result is obtained without additional training data, distinguishing AutoDev from LATS, which achieves 94.4\%. Furthermore, the AutoDev framework enhances GPT-4 performance from 67\% to \autodevCodeGenPassOne\%, marking a 30\% relative improvement.

These results underscore AutoDev's capability to significantly enhance the performance of LLMs in completing software engineering tasks. However, it's worth noting that AutoDev's effectiveness may entail multiple inference calls and steps, as we delve into further detail in $RQ_3$.


\begin{table}[t]
\small 
\centering
\begin{tabular}{@{}lcccc@{}}
\toprule
Approach              & Model & Pass@1 & \begin{tabular}[c]{@{}c@{}}Passing\\ Coverage\end{tabular} & \begin{tabular}[c]{@{}c@{}}Overall\\ Coverage\end{tabular} \\
\midrule
Human                  & - & 100     & \humanTestGenCoverage &  \humanTestGenCoverage\\ 
\textbf{AutoDev}           & GPT-4 & \autodevTestGenPassOne    &  \autodevTestGenCoverage & 88.8\\
zero-shot (baseline) & GPT-4 & 75 & 99.3 & 74\\ \bottomrule
\end{tabular}
\caption{Test Generation results on HumanEval. AutoDev enhances Pass@1 compared to baseline, generating tests achieving comparable coverage to human-written tests.}
\label{tab:rq2}
\end{table}

\subsection*{$RQ_2$: How effective is AutoDev in test generation task?}
Table \ref{tab:rq2} presents the outcomes for $RQ_2$, comparing AutoDev with the zero-shot GPT-4 (baseline) and human-written tests in the test generation task. Since OpenAI has not evaluated GPT-4 on the test generation task, we obtained the zero-shot GPT-4 results by invoking the inference with instruction prompting on the same GPT-4 model used for AutoDev.

AutoDev attains a Pass@1 score of \autodevTestGenPassOne\% on the HumanEval dataset modified for the test generation task, exhibiting a 17\% relative improvement over the baseline that utilizes the same GPT-4 model. The correct tests generated by AutoDev (included in Pass@1) achieve a robust 99.3\% coverage, comparable to the human-written tests' coverage of 99.4\%. Additionally, Table \ref{tab:rq2} reports the overall coverage across the entire dataset of test cases, considering incorrect or failing tests as lacking coverage. In this regard, AutoDev achieves an 88.8\% coverage over the complete dataset of focal methods.

These results affirm AutoDev's prowess in addressing diverse software engineering tasks.

\subsection*{$RQ_3$: How efficient is AutoDev in completing tasks?}

\begin{figure}[t]
    \centering
    \includegraphics[width=0.5\textwidth]{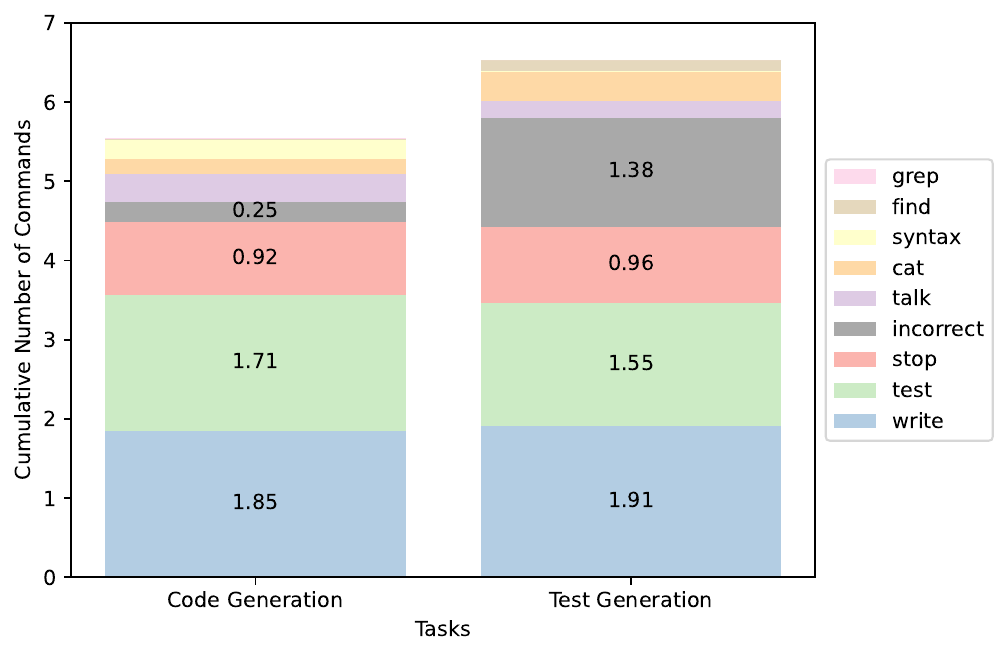}
    \caption{Cumulative number of commands used by AutoDev for an average task of Code and Test Generation}
    \label{fig:commands}
    
\end{figure}

Figure \ref{fig:commands} illustrates the cumulative number of commands used by AutoDev for both the Code Generation and Test Generation tasks, considering the average number of commands used for the evaluation of each HumanEval problem in $RQ_1$ and $RQ_2$.

For Code Generation, AutoDev executed an average of 5.5 commands, comprising 1.8 write operations, 1.7 test operations, 0.92 stop operations (indicating task completion), 0.25 incorrect commands, along with minimal retrieval (grep, find, cat), syntax check operations, and talk communication commands.

In the case of Test Generation, the average number of commands aligns with the Code Generation task. However, Test Generation involves more retrieval operations and an increased occurrence of incorrect operations, resulting in a total average of 6.5 commands for each run.

We classify a command as incorrect if it references an unavailable command or fails parsing (e.g., incorrect formatting or parameter count). The most prevalent incorrect commands involve AI agents mixing natural language with code or commands. Such issues could potentially be addressed through more flexible parsing or improved prompting, as discussed further in the Discussion section.

While AutoDev incurs more inference calls compared to approaches that generate candidate code in a single call, it is essential to note that AutoDev also performs testing and validation operations, tasks typically executed by developers to validate generated candidates. Testing and syntax operations invoked by AutoDev would be undertaken regardless by developers receiving code generated by AI, such as Copilot.

Furthermore, AutoDev often communicates task completion through \texttt{talk} commands, providing insights and interpretability of the solution. Another notable communication command contributing to the overall count is the \texttt{stop} command. This represents a relatively inexpensive inference call, generating only one token. Potential optimizations could involve batching such operations with other commands or inferences.

The average length of AutoDev conversations to solve each HumanEval problem in $RQ_1$ and $RQ_2$ is 1656 and 1863 tokens respectively. This encompasses the user's goal, messages from the AI agent, and responses from the Evaluation Environment. In comparison, the zero-shot GPT-4 (baseline) uses 200 tokens (estimated) for code generation and 373 tokens for test generation, on average, for each task. While AutoDev uses more tokens, a significant amount are spent for testing, validation, and explanation of its own generated code, going beyond what the baseline approach offers.

Finally, AutoDev incurs execution costs related to orchestrating AI Agents, managing conversations, and executing commands within a Docker environment. Notably, the Docker-based evaluation environment accounts for the major execution cost, exhibiting higher overhead compared to direct CLI commands within the user's environment. However, we designed AutoDev with security as a top priority, ensuring the secure execution and validation of AI-generated code.

\section{Discussion}

\subsection{AutoDev in Action}

Figures \ref{fig:example_1} (Part I) and \ref{fig:example_2} (Part II) provide a glimpse of AutoDev in action during a test generation task, with some interactions summarized or omitted for brevity. Initiating from Figure \ref{fig:example_1}, we set the goal for AutoDev to generate pytest test cases following a specific format. The AutoDev agent initiates the \texttt{write-new} command, providing the filepath and content of the test file. AutoDev executes the operation, confirming successful content writing to the agent. Subsequently, the AutoDev agent triggers the \texttt{test} operation, and AutoDev runs the test within its secure Docker environment, presenting the test execution report JSON (summarized in the figure for brevity).

\begin{figure}[H]
\vspace{-0.0cm}
    \centering
\begin{adjustbox}{width=0.5\textwidth}
\begin{tabular}{c}
\toprule
  User - Objective\\
\midrule
\begin{minipage}[t]{0.45\textwidth}
\begin{minted}{text}
Write a Pytest test with 4-5 assertions that covers the following
function, located at human_answer.py:
\end{minted}
\begin{minted}[escapeinside=||]{python}
def is_bored(S):
    """
    You'll be given a string of words, and your task is to count 
    the number of boredoms. A boredom is a sentence that starts 
    with the word "I". Sentences are delimited by '.', '?' or '!'.
   
    For example:
    >>> is_bored("Hello world")
    0
    >>> is_bored("The sky is blue. The sun is shining. 
    I love this weather")
    1
    """

    import re
    sentences = re.split(r'[.?!]\s*', S)
    return sum(sentence[0:2] == 'I ' for sentence in sentences)
\end{minted}
\begin{minted}{text}
Write the test to a new file /HumanEval_91/test_HumanEval_91.py. 
The test method should be written in the following format:
\end{minted}
\begin{minted}[escapeinside=||]{python}
from .human_answer import *
def test_function_name():
    # assertions for the given function here
\end{minted}
\begin{minted}{text}
Run your generated test and confirm it passes.
\end{minted}
\end{minipage}\\
\midrule
AutoDev - Agent \\
\midrule
\begin{minipage}[t]{0.45\textwidth}
\begin{minted}{text}
write-new /HumanEval_91/test_HumanEval_91.py
\end{minted}
\begin{minted}[breaklines, escapeinside=||]{python}
from .human_answer import is_bored
import pytest

def test_is_bored():
    assert is_bored('Hello world') == 0
    assert is_bored('I am bored. This is boring!') == 2
    assert is_bored('The sky is blue. The sun is shining. I love this weather.') == 1
    assert is_bored('I think, therefore I am. I am bored?') == 2
    assert is_bored('') == 0
\end{minted}
\end{minipage}\\
\midrule
AutoDev - Environment \\
\midrule
\begin{minipage}[t]{0.45\textwidth}
\begin{minted}{text}
Content successfully written to /HumanEval_91/test_HumanEval_91.py
\end{minted}
\end{minipage}\\
\midrule
AutoDev - Agent \\
\midrule
\begin{minipage}[t]{0.45\textwidth}
\begin{minted}{text}
test
\end{minted}
\end{minipage}\\
\midrule
AutoDev - Environment \\
\midrule
\begin{minipage}[t]{0.45\textwidth}
\begin{minted}[breaklines]{text}
...
'summary': {'failed': 1, 'total': 1, 'collected': 1}
...
'message': "AssertionError: assert 1 == 2\n +  where 1 = is_bored('I am bored. This is boring!')
...
\end{minted}
\end{minipage}\\
\midrule
AutoDev - Agent \\
\midrule
\begin{minipage}[t]{0.45\textwidth}
\begin{minted}[breaklines]{text}
The test case that failed expected `is_bored('I am bored. This is boring!')` to return `2`, but according to the function implementation, it should return `1` because the function counts sentences that start with the word "I" followed by a space. The sentence "This is boring!" does not start with "I", so it should not be counted.

The test case is incorrect and needs to be fixed. Let's correct the test case to match the expected behavior of the function.
\end{minted}

\end{minipage} \\
\\
\bottomrule
\end{tabular}
\end{adjustbox}
\vspace{-0.0cm}
\caption{AutoDev in Test Generation scenario (part I)}
\vspace{-0.0cm}
\label{fig:example_1}
\end{figure}

\begin{figure}[h]
\vspace{-0.0cm}
    \centering
\begin{adjustbox}{width=0.5\textwidth}
\begin{tabular}{c}
\toprule
  AutoDev - Agent\\
\midrule
\begin{minipage}[t]{0.45\textwidth}
\begin{minted}{text}
write /HumanEval_91/test_HumanEval_91.py 5-5
\end{minted}
\begin{minted}[escapeinside=||]{python}
"    assert is_bored('I am bored. This is boring!') == 1"
\end{minted}
\end{minipage}\\
\midrule
AutoDev - Environment \\
\midrule
\begin{minipage}[t]{0.45\textwidth}
\begin{minted}{text}
File updates successfully.
\end{minted}
\end{minipage}\\
\midrule
AutoDev - Agent \\
\midrule
\begin{minipage}[t]{0.45\textwidth}
\begin{minted}{text}
test
\end{minted}
\end{minipage}\\
\midrule
AutoDev - Environment \\
\midrule
\begin{minipage}[t]{0.45\textwidth}
\begin{minted}{text}
'summary': {'passed': 1, 'total': 1, 'collected': 1}
\end{minted}
\end{minipage}\\
\midrule
AutoDev - Agent \\
\midrule
\begin{minipage}[t]{0.45\textwidth}
\begin{minted}[breaklines]{text}
The test suite has passed successfully. All tests for the `is_bored` function are now correct and the function behaves as expected. The goal has been reached. 
\end{minted}
\end{minipage} \\
\\
\bottomrule
\end{tabular}
\end{adjustbox}
\vspace{-0.0cm}
\caption{AutoDev in Test Generation scenario (part II)}
\vspace{-0.0cm}
\label{fig:example_2}
\end{figure}

The AutoDev agent identifies an error in the pytest output, recognizing that a fix is required to align the test with the expected behavior of the function.

Continuing in Figure \ref{fig:example_2}, the AutoDev agent issues the \texttt{write} command, specifying the filepath and line number range (5-5) for rewriting the incorrect assert statement. Following this, the AutoDev agent proceeds with the test execution, which now succeeds. Recognizing the completion of the goal, the AutoDev agent concludes the conversation.

This example highlights AutoDev's capability to self-evaluate its generated code and address bugs within its own output. Additionally, it demonstrates how AutoDev facilitates user insight into agent actions, allowing agents to communicate during the task.

\subsection{Multi-Agent Collaboration}
AutoDev supports multi-agent collaboration on tasks, orchestrated by the Agent Scheduler. In our evaluation, given the relative simplicity of the HumanEval dataset, we limited our setup to a single GPT-4 agent for the tasks.

However, preliminary results indicate the positive impact of multi-agent collaboration on more complex tasks. Experimenting with an AI Developer and AI Reviewer, each with distinct responsibilities and available actions, collaborating on fixing a complex bug revealed interesting interactions. The AI Reviewer could pre-emptively identify AI Developer mistakes before code validation actions were executed and provide relevant suggestions.

Our future plans involve expanding evaluations to incorporate more complex scenarios where multi-agent collaboration can significantly enhance AutoDev's performance.

\subsection{Human in the Loop}
AutoDev allows AI agents to communicate progress on tasks or request human feedback using the \texttt{talk} and \texttt{ask} commands, respectively. Anecdotally, these commands have proven helpful for developers using AutoDev to understand the agent's intentions and gain insights into the agent's plan. The addition of the \texttt{ask} command was in direct response to a developer's request during our pilot study, where they wanted the ability to provide feedback when the agents appeared uncertain about the next actions.

Our future plans involve deeper integration of humans within the AutoDev loop, allowing users to interrupt agents and provide prompt feedback.

\subsection{AutoDev Integrations}
Our pilot study involved developers using AutoDev as a CLI command, with the conversation available for observation within the VSCode IDE.

Moving forward, our goal is to integrate AutoDev into IDEs, creating a chatbot experience, and incorporate it into CI/CD pipelines and PR review platforms. We envision developers assigning tasks and issues to AutoDev, reviewing results within a PR system, and further streamlining the software development workflow.

\section{Related Work}
Our work builds upon an extensive body of literature that applies AI to various Software Engineering tasks. In this section, we explore recent developments and contextualize AutoDev within this rich research landscape.

\subsection{AI in Software Engineering}

The integration of AI, particularly Large Language Models (LLMs), into software engineering has witnessed substantial progress. Models like GPT-3 \cite{floridi2020gpt}, InstructGPT \cite{ouyang2022training}, and GPT-4 \cite{openai2023gpt4} have leveraged the Transformer architecture \cite{vaswani2017attention} to comprehend and generate not only natural language, but also source code. The massive parameter sizes of LLMs, such as those in Gropher \cite{rae2022scaling} and Megatron-turing NLG \cite{smith2022using}, and GPT-4 \cite{openai2023gpt4} allowed these AI models to achieve impressive performance across diverse tasks. 

As software development practices continue to evolve, the integration of cutting-edge technologies becomes paramount for enhancing developer productivity \cite{chen2021codex}. Among the notable advancements, the utilization of LLMs within Integrated Development Environments (IDEs) has garnered significant attention \cite{nam2023inide, vscuda}.
LLMs, including prominent models such as OpenAI's GPT-3.5 \cite{openaigpt35} and GPT-4 \cite{openai2023gpt4}, as well as robust open-source models like Code Llama \cite{roziere2023code}, exhibit the potential to act as intelligent programming assistants. 

In this paper, we introduce AutoDev, a comprehensive framework for autonomous software engineering tasks within a secure development environment. AutoDev extends beyond existing works by providing a versatile tools library, empowering AI agents to autonomously perform intricate tasks, such as code editing, testing, and integration. AutoDev is also LLM-agnostic, with an infrastructure that allows a diverse set of AI models, with different parameter size and architectures, to collaborate on a given task.

\subsection{Evaluation of LLMs in Software Engineering}

Evaluating LLMs for software engineering tasks poses unique challenges. Traditional language-based metrics, such as BLEU, have been the focus of previous research, with evaluations conducted on static datasets like GLUE \cite{wang2019glue} and BIGBench \cite{srivastava2023imitation}. However, these metrics often fall short in capturing essential programming aspects like syntax correctness and execution-based metrics such as build and testing.

CodeXGLUE \cite{lu2021codexglue} has addressed these limitations by providing a comprehensive evaluation platform for LLMs in software engineering. It offers a diverse benchmark dataset along with baseline models like CodeBERT and CodeGPT.

HumanEval \cite{chen2021codex} contributes to the field by focusing on the functional correctness of LLMs, introducing a benchmark dataset of hand-written programming problems in Python.

In recent developments, the Copilot Evaluation Harness \cite{agarwal2024copilot} builds upon prior works in the literature, aiming to enhance their contributions. Similar to HumanEval,  the Copilot Evaluation Harness incorporates considerations of code execution, but extending the spectrum of software engineering tasks (code, test, and documentation generation, workspace understanding and query resolution) as well as increasing the metrics used for evaluation. This evaluation harness also encompasses large and real-world codebases. 

While our current evaluation relies on HumanEval for assessing the effectiveness of AutoDev in two coding tasks, our future work aims to extend this assessment to more challenging and real-world datasets, such as those offered by the Copilot Evaluation Harness.

\subsection{AI in Software Engineering Interactions}

While prior works have explored the intersection of AI and software engineering, few have delved into AI-guided programming within IDE interactions. AutoDev, as introduced in this paper, draws inspiration from existing works in the literature while enhancing their contributions. Notable examples include Auto-GPT\cite{autogpt}, LATS (Language Agent Tree Search)\cite{zhou2023language}, and Reflexion\cite{shinn2023reflexion}, each presenting unique approaches to AI-driven tasks.

Auto-GPT\cite{autogpt} operates by pairing GPT-3.5 and GPT-4 with a companion bot, enabling users to instruct these language models on specific goals. The companion bot utilizes GPT-3.5 and GPT-4, along with various programs, to execute the necessary steps for goal achievement.

LATS \cite{zhou2023language} , on the other hand, is a general framework that synergizes the capabilities of LLMs in planning, acting, and reasoning. Inspired by Monte Carlo tree search, commonly used in model-based reinforcement learning, LATS employs LLMs as agents, value functions, and optimizers, enhancing decision-making. It introduces an environment for external feedback, offering a deliberate and adaptive problem-solving mechanism.

Reflexion\cite{shinn2023reflexion} introduces a novel framework to reinforce language agents through linguistic feedback. Reflexion agents verbally reflect on task feedback signals, maintaining their reflective text in an episodic memory buffer for improved decision-making. This flexible framework incorporates various types and sources of feedback signals and exhibits significant improvements over baseline agents in diverse tasks, including sequential decision-making, coding, and language reasoning. 

AutoDev specializes these ideas for the Software Engineering realm, offering a flexible framework that allows AI agents to complete complex SE tasks in full autonomy. Our work aims to bridge the gap between traditional software engineering practices and AI-driven automation, facilitating collaborative efforts between developers and AI agents. By introducing a versatile tools library, AutoDev empowers AI agents to autonomously perform intricate tasks, providing a promising advancement in the landscape of AI-assisted software development.

\section{Conclusion}
In this paper, we introduced AutoDev, a framework enabling AI Agents to autonomously interact with repositories, perform actions, and tackle complex software engineering tasks. We've shifted the responsibility of extracting relevant context for software engineering tasks and validating AI-generated code from users (mainly developers) to the AI agents themselves. Agents are now empowered to retrieve context through Retrieval actions and validate their code generation through Build, Execution, Testing, and Validation actions.

The developer's role within the AutoDev framework transforms from manual actions and validation of AI suggestions to a supervisor overseeing multi-agent collaboration on tasks, with the option to provide feedback. Developers can monitor AutoDev's progress toward goals by observing the ongoing conversation used for communication among agents and the repository.

Our evaluation on the HumanEval dataset for code and test generation showcased impressive results, achieving a Pass@1 score of 91.5 for code generation—a second-best result on the leaderboard at the time of writing, and the best among approaches requiring no extra training data. AutoDev also excelled in test generation with a Pass@1 score of \autodevTestGenPassOne\%, achieving a \autodevTestGenCoverage\% coverage from passing tests.

Looking ahead, our goal for future work is to integrate AutoDev into IDEs as a chatbot experience and incorporate it into CI/CD pipelines and PR review platforms.

\bibliographystyle{acm}
\bibliography{main}

\begin{thebibliography}{10}

\bibitem{paperswithcode-humaneval}
Code generation on humaneval - state-of-the-art.
\newblock \url{https://paperswithcode.com/sota/code-generation-on-humaneval}, 2024.
\newblock Accessed: 2024-02-27.

\bibitem{copilot}
Github copilot: Your ai pair programmer.
\newblock \url{https://github.com/features/copilot}, 2024.

\bibitem{agarwal2024copilot}
{\sc Agarwal, A., Chan, A., Chandel, S., Jang, J., Miller, S., Moghaddam, R.~Z., Mohylevskyy, Y., Sundaresan, N., and Tufano, M.}
\newblock Copilot evaluation harness: Evaluating llm-guided software programming, 2024.

\bibitem{vscuda}
{\sc Chen, B., Mustakin, N., Hoang, A., Fuad, S., and Wong, D.}
\newblock Vscuda: Llm based cuda extension for visual studio code.
\newblock In {\em Proceedings of the SC '23 Workshops of The International Conference on High Performance Computing, Network, Storage, and Analysis\/} (New York, NY, USA, 2023), SC-W '23, Association for Computing Machinery, p.~11–17.

\bibitem{chen2021codex}
{\sc Chen, M., Tworek, J., Jun, H., Yuan, Q., de~Oliveira~Pinto, H.~P., et~al.}
\newblock Evaluating large language models trained on code.

\bibitem{ding2024crosscodeeval}
{\sc Ding, Y., Wang, Z., Ahmad, W., Ding, H., Tan, M., Jain, N., Ramanathan, M.~K., Nallapati, R., Bhatia, P., Roth, D., et~al.}
\newblock Crosscodeeval: A diverse and multilingual benchmark for cross-file code completion.
\newblock {\em Advances in Neural Information Processing Systems 36\/} (2024).

\bibitem{floridi2020gpt}
{\sc Floridi, L., and Chiriatti, M.}
\newblock Gpt-3: Its nature, scope, limits, and consequences.
\newblock {\em Minds and Machines 30\/} (2020), 681--694.

\bibitem{autogpt}
{\sc Gravitas, S.}
\newblock Autogpt.
\newblock \url{https://github.com/Significant-Gravitas/AutoGPT}, 2024.
\newblock GitHub repository.

\bibitem{lu2021codexglue}
{\sc Lu, S., Guo, D., Ren, S., Huang, J., Svyatkovskiy, A., Blanco, A., Clement, C., Drain, D., Jiang, D., Tang, D., Li, G., Zhou, L., Shou, L., Zhou, L., Tufano, M., Gong, M., Zhou, M., Duan, N., Sundaresan, N., Deng, S.~K., Fu, S., and Liu, S.}
\newblock Codexglue: A machine learning benchmark dataset for code understanding and generation, 2021.

\bibitem{nam2023inide}
{\sc Nam, D., Macvean, A., Hellendoorn, V., Vasilescu, B., and Myers, B.}
\newblock In-ide generation-based information support with a large language model, 2023.

\bibitem{openai2024gpt4}
{\sc OpenAI, :, Achiam, J., Adler, S., Agarwal, S., Ahmad, L., Akkaya, I., Aleman, F.~L., Almeida, D., Altenschmidt, J., Altman, S., Anadkat, S., Avila, R., Babuschkin, I., Balaji, S., Balcom, V., Baltescu, P., Bao, H., Bavarian, M., Belgum, J., Bello, I., Berdine, J., Bernadett-Shapiro, G., Berner, C., Bogdonoff, L., Boiko, O., Boyd, M., Brakman, A.-L., Brockman, G., Brooks, T., Brundage, M., Button, K., Cai, T., Campbell, R., Cann, A., Carey, B., Carlson, C., Carmichael, R., Chan, B., Chang, C., Chantzis, F., Chen, D., Chen, S., Chen, R., Chen, J., Chen, M., Chess, B., Cho, C., Chu, C., Chung, H.~W., Cummings, D., Currier, J., Dai, Y., Decareaux, C., Degry, T., Deutsch, N., Deville, D., Dhar, A., Dohan, D., Dowling, S., Dunning, S., Ecoffet, A., Eleti, A., Eloundou, T., Farhi, D., Fedus, L., Felix, N., Fishman, S.~P., Forte, J., Fulford, I., Gao, L., Georges, E., Gibson, C., Goel, V., Gogineni, T., Goh, G., Gontijo-Lopes, R., Gordon, J., Grafstein, M., Gray, S., Greene, R., Gross, J., Gu, S.~S., Guo, Y.,
  Hallacy, C., Han, J., Harris, J., He, Y., Heaton, M., Heidecke, J., Hesse, C., Hickey, A., Hickey, W., Hoeschele, P., Houghton, B., Hsu, K., Hu, S., Hu, X., Huizinga, J., Jain, S., Jain, S., Jang, J., Jiang, A., Jiang, R., Jin, H., Jin, D., Jomoto, S., Jonn, B., Jun, H., Kaftan, T., Łukasz Kaiser, Kamali, A., Kanitscheider, I., Keskar, N.~S., Khan, T., Kilpatrick, L., Kim, J.~W., Kim, C., Kim, Y., Kirchner, J.~H., Kiros, J., Knight, M., Kokotajlo, D., Łukasz Kondraciuk, Kondrich, A., Konstantinidis, A., Kosic, K., Krueger, G., Kuo, V., Lampe, M., Lan, I., Lee, T., Leike, J., Leung, J., Levy, D., Li, C.~M., Lim, R., Lin, M., Lin, S., Litwin, M., Lopez, T., Lowe, R., Lue, P., Makanju, A., Malfacini, K., Manning, S., Markov, T., Markovski, Y., Martin, B., Mayer, K., Mayne, A., McGrew, B., McKinney, S.~M., McLeavey, C., McMillan, P., McNeil, J., Medina, D., Mehta, A., Menick, J., Metz, L., Mishchenko, A., Mishkin, P., Monaco, V., Morikawa, E., Mossing, D., Mu, T., Murati, M., Murk, O., Mély, D., Nair, A.,
  Nakano, R., Nayak, R., Neelakantan, A., Ngo, R., Noh, H., Ouyang, L., O'Keefe, C., Pachocki, J., Paino, A., Palermo, J., Pantuliano, A., Parascandolo, G., Parish, J., Parparita, E., Passos, A., Pavlov, M., Peng, A., Perelman, A., de~Avila Belbute~Peres, F., Petrov, M., de~Oliveira~Pinto, H.~P., Michael, Pokorny, Pokrass, M., Pong, V.~H., Powell, T., Power, A., Power, B., Proehl, E., Puri, R., Radford, A., Rae, J., Ramesh, A., Raymond, C., Real, F., Rimbach, K., Ross, C., Rotsted, B., Roussez, H., Ryder, N., Saltarelli, M., Sanders, T., Santurkar, S., Sastry, G., Schmidt, H., Schnurr, D., Schulman, J., Selsam, D., Sheppard, K., Sherbakov, T., Shieh, J., Shoker, S., Shyam, P., Sidor, S., Sigler, E., Simens, M., Sitkin, J., Slama, K., Sohl, I., Sokolowsky, B., Song, Y., Staudacher, N., Such, F.~P., Summers, N., Sutskever, I., Tang, J., Tezak, N., Thompson, M.~B., Tillet, P., Tootoonchian, A., Tseng, E., Tuggle, P., Turley, N., Tworek, J., Uribe, J. F.~C., Vallone, A., Vijayvergiya, A., Voss, C., Wainwright,
  C., Wang, J.~J., Wang, A., Wang, B., Ward, J., Wei, J., Weinmann, C., Welihinda, A., Welinder, P., Weng, J., Weng, L., Wiethoff, M., Willner, D., Winter, C., Wolrich, S., Wong, H., Workman, L., Wu, S., Wu, J., Wu, M., Xiao, K., Xu, T., Yoo, S., Yu, K., Yuan, Q., Zaremba, W., Zellers, R., Zhang, C., Zhang, M., Zhao, S., Zheng, T., Zhuang, J., Zhuk, W., and Zoph, B.}
\newblock Gpt-4 technical report, 2024.

\bibitem{openaigpt35}
{\sc OpenAI}.
\newblock Gpt 3.5 models, 2023.

\bibitem{openai2023gpt4}
{\sc OpenAI}.
\newblock Gpt-4 technical report, 2023.

\bibitem{ouyang2022training}
{\sc Ouyang, L., Wu, J., Jiang, X., Almeida, D., Wainwright, C., Mishkin, P., Zhang, C., Agarwal, S., Slama, K., Ray, A., et~al.}
\newblock Training language models to follow instructions with human feedback.
\newblock {\em Advances in Neural Information Processing Systems 35\/} (2022), 27730--27744.

\bibitem{rae2022scaling}
{\sc Rae, J.~W., Borgeaud, S., Cai, T., Millican, K., and Others}.
\newblock Scaling language models: Methods, analysis \& insights from training gopher, 2022.

\bibitem{roziere2023code}
{\sc Roziere, B., Gehring, J., Gloeckle, F., Sootla, S., Gat, I., Tan, X.~E., Adi, Y., Liu, J., Remez, T., Rapin, J., et~al.}
\newblock Code llama: Open foundation models for code.
\newblock {\em arXiv preprint arXiv:2308.12950\/} (2023).

\bibitem{shinn2023reflexion}
{\sc Shinn, N., Cassano, F., Berman, E., Gopinath, A., Narasimhan, K., and Yao, S.}
\newblock Reflexion: Language agents with verbal reinforcement learning, 2023.

\bibitem{smith2022using}
{\sc Smith, S., Patwary, M., Norick, B., LeGresley, P., Rajbhandari, S., Casper, J., Liu, Z., Prabhumoye, S., Zerveas, G., Korthikanti, V., Zhang, E., Child, R., Aminabadi, R.~Y., Bernauer, J., Song, X., Shoeybi, M., He, Y., Houston, M., Tiwary, S., and Catanzaro, B.}
\newblock Using deepspeed and megatron to train megatron-turing nlg 530b, a large-scale generative language model, 2022.

\bibitem{srivastava2023imitation}
{\sc Srivastava, A., Rastogi, A., Rao, A., Shoeb, A. A.~M., et~al.}
\newblock Beyond the imitation game: Quantifying and extrapolating the capabilities of language models, 2023.

\bibitem{vaswani2017attention}
{\sc Vaswani, A., Shazeer, N., Parmar, N., Uszkoreit, J., Jones, L., Gomez, A.~N., Kaiser, {\L}., and Polosukhin, I.}
\newblock Attention is all you need.
\newblock {\em Advances in neural information processing systems 30\/} (2017).

\bibitem{wang2019glue}
{\sc Wang, A., Singh, A., Michael, J., Hill, F., Levy, O., and Bowman, S.~R.}
\newblock Glue: A multi-task benchmark and analysis platform for natural language understanding, 2019.

\bibitem{wu2023autogen}
{\sc Wu, Q., Bansal, G., Zhang, J., Wu, Y., Li, B., Zhu, E., Jiang, L., Zhang, X., Zhang, S., Liu, J., Awadallah, A.~H., White, R.~W., Burger, D., and Wang, C.}
\newblock Autogen: Enabling next-gen llm applications via multi-agent conversation, 2023.

\bibitem{zhou2023language}
{\sc Zhou, A., Yan, K., Shlapentokh-Rothman, M., Wang, H., and Wang, Y.-X.}
\newblock Language agent tree search unifies reasoning acting and planning in language models, 2023.

\end{thebibliography}

\end{document}